# India's residential space cooling transition: Decarbonization ambitions since the turn of millennium☆

Ran Yan [1], Nan Zhou [3], Minda Ma [2*, 3], Chao Mao [1*]

1. School of Management Science and Real Estate, Chongqing University, Chongqing, 400045, PR China
2. School of Architecture and Urban Planning, Chongqing University, Chongqing, 400045, PR China
3. Building Technology and Urban Systems Division, Energy Technologies Area, Lawrence Berkeley National Laboratory, Berkeley, CA 94720, United States

- The leading corresponding author: Dr. Minda Ma, Email: maminda@lbl.gov
  Homepage: https://buildings.lbl.gov/people/minda-ma
  https://chongjian.cqu.edu.cn/info/1556/6706.htm

---

☆ The coauthors from Lawrence Berkeley National Laboratory declare that this manuscript was authored by an author at Lawrence Berkeley National Laboratory under Contract No. DE-AC02-05CH11231 with the U.S. Department of Energy. The U.S. Government retains, and the publisher, by accepting the article for publication, acknowledges, that the U.S. Government retains a non-exclusive, paid-up, irrevocable, world-wide license to publish or reproduce the published form of this manuscript, or allows others to do so, for U.S. Government purposes.

## Highlights

- Indian space cooling carbon intensity reached 514 $kgCO_2$/household in 2022, up 6% annually since 2000.
- Income per capita was the most positive contributor to the increase in carbon intensity in the last 22 years.
- Fans contributed to >70% of carbon intensity of residential space cooling in urban and rural India in 2022.
- States with higher decarbonization potential were characterized by either high income or hot climates.
- Promoting energy-efficient building designs can be prioritized to achieve affordable space cooling.

## Abstract

As an emerging emitter poised for significant growth in space cooling demand, India requires comprehensive insights into historical emission trends and decarbonization performance to shape future low-carbon cooling strategies. By integrating a bottom-up demand resource energy analysis model and a top-down decomposition method, this study is the first to conduct a state-level analysis of carbon emission trends and the corresponding decarbonization efforts for residential space cooling in urban and rural India from 2000 to 2022. The results indicate that (1) the carbon intensity of residential space cooling in India increased by 292.4% from 2000 to 2022, reaching 513.8 kilograms of carbon dioxide per household. The net state domestic product per capita, representing income, emerged as the primary positive contributor. (2) The increase in carbon emissions from space cooling can be primarily attributed to the use of fans. While fan-based space cooling has nearly saturated Indian urban households, it is anticipated to persist as the primary cooling method in rural households for decades. (3) States with higher decarbonization potential are concentrated in two categories: those with high household income and substantial cooling appliance ownership and those with pronounced unmet cooling demand but low household income and hot climates. Furthermore, it is believed that promoting energy-efficient building designs can be prioritized to achieve affordable space cooling. Overall, this study serves as an effective foundation for formulating and promoting India's future cooling action plan, addressing the country's rising residential cooling demands and striving toward its net-zero goal by 2070.

**Abbreviation notation**

ACs – Air conditioners

$CO_2$ – carbon dioxide

DREAM – Demand Resource Energy Analysis Model

DSD – Decomposing structural decomposition

ECBC – Energy Conservation Building Code

ICAP – India Cooling Action Plan

IEA – International Energy Agency

INR – Indian Rupee

ISEER – Indian Seasonal Energy Efficient Ratio

kg$CO_2$ – Kilograms of carbon dioxide

kWh – kilowatt hours

LEAP – Low Emissions Analysis Platform

LED – light-emitting diodes

$m^2$ – Square meters

Mt$CO_2$ – Megatons of carbon dioxide

NSDP – Net State Domestic Product

S&L – Standards and Labeling

W – watt

**Nomenclature**

$C$ – Carbon emissions from residential space cooling

$c$ – Carbon intensity (carbon emissions per household) associated with residential space cooling

$c_a$ – Carbon intensity for fans in Indian residential space cooling

$c_f$ – Carbon intensity for fans in Indian residential space cooling

$c_r$ – Carbon intensity for room air conditioners (ACs) in Indian residential space cooling

d$F$ – The slack component introduced in the DSD method

d$F_i$ $(i = 1, 2, 3)$ – The shift component introduced in the DSD method

$\Delta DC$ – Total decarbonization

$\Delta Dc$ – Decarbonization intensity

$\Delta d$ – Decarbonization efficiency

$E$ – Energy consumption associated with residential space cooling

$E_i$ $(i = 1,2,3)$ – Energy consumption associated with residential space cooling appliance $i$

$e$ – Energy intensity (energy consumption per unit NSDP)

$e_i$ $(i = 1,2,3)$ – Energy intensity associated with residential space cooling appliance $i$

$f_r$ – Room floorspace supplied with room AC services per household

$H$ – Amount of households

$k$ – Emission factors

$k_i$ – Emission factor of the energy consumption from appliance $i$

$l_r$ – Annual cooling load per floorspace

$N$ – Net State Domestic Product (NSDP)

$n$ – NSDP per capita

$P$ – Population size

$p$ – Population per household

$t_a$ – Annual usage hours of air coolers

$t_f$ – Annual usage hours of fans

$u_a$ – Number of air coolers being used per household

$u_f$ – Number of fans being used per household

$w_i$ – Share of $E_i$ in $E$ (representing energy structure of the three main appliances)

$x$ – Locale type (urban or rural)

$y$ – Type of appliances (room ACs, fans, and air coolers)

$\lambda_a$ – Average rated power of air coolers commonly used in Indian households

$\lambda_f$ – Power of fans at 210 $m^3$/min air delivery

$\lambda_r$ – ISEER

## 1. Introduction

### *1.1. Background*

Climate warming is expected to drive a significant increase in cooling demand worldwide, particularly in rapidly urbanizing economies like India [1], which faces challenges such as extreme heat waves [2]. The International Energy Agency (IEA) projects that by 2050, India's residential cooling electricity demand will surpass the current total electricity consumption of Africa [3], complicating the nation's goal of achieving net-zero emissions by 2070. Promoting energy-efficient practices is both costly and time-consuming, impacting 90% of India's population [4]. Additionally, India lacks regular surveys on residential energy use, making it difficult to track cooling activities across climate zones and income classes. Gaps in research on carbon emissions and decarbonization efforts at the state level may lead to inaccurate predictions of future energy use, hindering effective policy development [5].

### *1.2. Literature review*

Research by the IEA has highlighted India's limited access to cooling, with per capita energy consumption for space cooling at just 69 kWh—only a quarter of the global average [6]. By 2021, only 12.6% of urban households and 1.2% of rural households in India had air conditioners (ACs), with most people relying primarily on fans [7]. However, as heat waves intensify, fans are becoming insufficient for thermal comfort. Economic and population growth are driving rapid increases in cooling demand, which is projected to significantly affect future electricity consumption [8].

Several studies have undertaken modeling efforts, incorporating various climate scenarios, to predict India's space cooling demand through 2050 [9, 10]. These analyses typically included factors like cooling degree days [11], socioeconomic advancements [12], and population growth [13], emphasizing the importance of energy-efficient technologies [14] and improved cooling systems [15, 16]. However, earlier predictions often overestimated the growth of AC use, given India's economic and grid constraints [17]. Income has emerged as a significant driver of AC adoption, as studies like those by Davis et al. [18] show the rising demand for cooling, especially among India's growing middle class. While some studies have explored cooling patterns in Indian urban households [19, 20], rural areas and lower-income groups remain underrepresented.

In regard to modeling methods for residential space cooling, most studies used bottom-up modeling approaches to analyze energy consumption and carbon emissions [21, 22]. Classic tools like the low emissions analysis platform (LEAP), developed by the Stockholm Environment Institute, were widely used [23]. Furthermore, the decomposing structural decomposition (DSD) method [24] has proven effective in analyzing changes in carbon emissions from residential space cooling, breaking down factors affecting emissions and highlighting decarbonization potential by matching different appliances with appropriate energy management.

### *1.3. Motivation, contributions, and the organization*

Based on the review of existing studies, a comprehensive analysis of historical carbon emissions from residential space cooling in India is essential. This analysis must address two critical gaps: First, the study should examine both urban and rural households across Indian states, accounting for regional disparities that are often overlooked in national-level studies [8, 10]. It should also consider recent trends, such as COVID-19, extreme heat events, and national cooling plans [15, 25]. Understanding cooling appliance usage patterns and energy consumption across regions is crucial, especially in the context of climate change and socioeconomic shifts. Second, the study should investigate the drivers of decarbonization in space cooling. Although the India Cooling Action Plan (ICAP) aims to reduce energy use by 25-40%, the residential sector lacks detailed strategies [26]. By applying decomposition methods, the study can identify regions with greater decarbonization potential and highlight which appliances offer the most significant short- to medium-term benefits. Addressing these research gaps is crucial to avoid inaccurate projections of future energy consumption and carbon emissions from residential space cooling in India, which could hinder the government's ability to prioritize cooling policies within budget and time constraints. Three critical research questions arise from these gaps, requiring urgent attention:

- What are the historical features of space cooling emissions in India's urban and rural households?

- How can decarbonization efforts from residential space cooling among Indian states be measured?
- Which solutions will expedite the future decarbonization of residential space cooling in India?

This study addresses the above challenges by integrating the bottom-up demand resource energy analysis model (DREAM) and the top-down DSD method for the first time. This work begins with a microlevel investigation of residential space cooling activity, examining the historical evolution of carbon intensity for space cooling from 2000 to 2022 in urban and rural households across Indian states—addressing the first question above. Following this, a macrolevel exploration delves into societal, economic, climatic, and technological factors influencing changes in carbon intensity for India's residential space cooling and further evaluates the historical performance of Indian states in decarbonizing urban and rural residential space cooling across four emission scales—addressing the second question. Finally, the study assesses the effectiveness of initiated residential space cooling policies, seeking optimal paths to expedite future decarbonization—addressing the third question.

**The most significant contribution of this study** lies in its comprehensive exploration of 33 states and union territories in India, distinguishing between urban and rural areas and focusing on primary residential space cooling appliances. The analysis reviews the changes in carbon intensity and the decarbonization performance of residential space cooling from 2000 to 2022, addressing a gap in previous studies. By thoroughly examining the heterogeneity in the economy, climate, and consumption behavior reflected in carbon emissions from residential space cooling in urban and rural areas across various states, this study enables the prediction of future trends in space cooling appliance ownership and the identification of regions with high decarbonization potential. Given the substantial budgets required for deploying key technologies in decarbonization and the urgent goal of achieving carbon neutrality by 2070, these insights provide a nuanced and effective basis for formulating and promoting future cooling policies. This involves identifying the priority decarbonization strategies for space cooling appliances and determining the states to target for future promotion—a critical strategic consideration for the Indian government.

The rest of this study unfolds as follows: Section 2 introduces the materials and methods, which include the bottom-up DREAM and the top-down DSD methods, along with the data sources. Section 3 illustrates the historical evolution and related contributors to space cooling carbon intensity in Indian households. Section 4 examines historical decarbonization performance across Indian states and discusses corresponding decarbonization strategy trajectories. Finally, Section 5 summarizes the key findings and proposes future related work.

## 2. Materials and methods

This study employed an integration of a bottom-up emissions model and a top-down decomposition model to conduct a comprehensive examination of historical carbon emissions from residential space cooling in India. To offer an intuitive overview of the logic of this study, Fig. 1 illustrates the relationships among the methods, the variables utilized in this article, the results obtained, and further discussion.

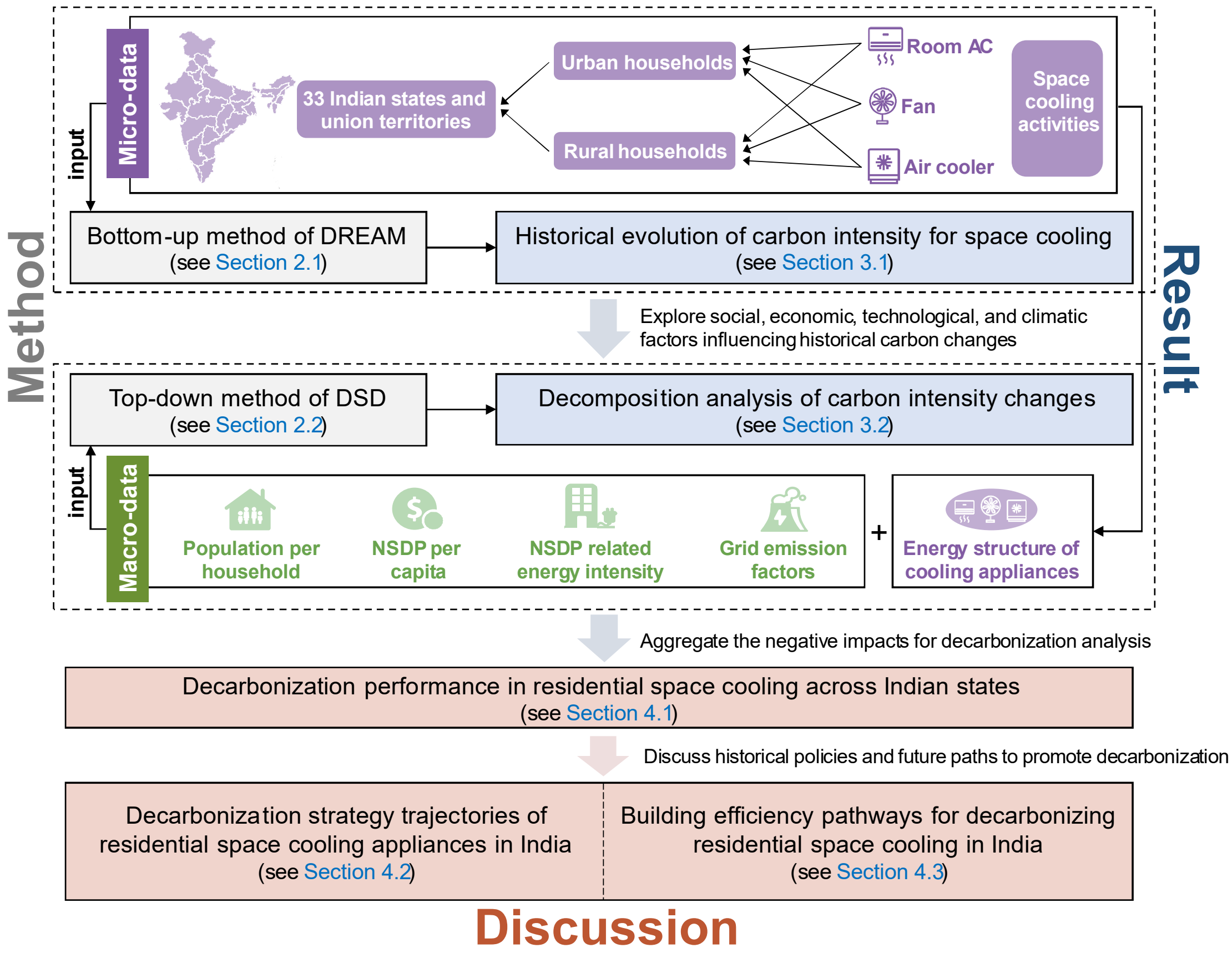

**Fig. 1.** Framework of this study.

### *2.1. Bottom-up DREAM model of residential space cooling*

This study utilized the modeling approach of the India DREAM [10], developed by the Lawrence

Berkeley National Laboratory, to analyze household energy consumption. DREAM employs a bottom-up methodology, offering a comprehensive quantitative energy assessment by systematically accounting for the energy consumption of individual appliances across various end-uses within buildings. Subsequently, different sources of energy consumption are multiplied by corresponding carbon emission factors to estimate the total carbon emissions from building operations. This step can be facilitated by the LEAP developed by the Stockholm Environment Institute. Key considerations in the modeling process include the energy intensity of end-uses, penetration rates of various household appliances, the energy efficiency of appliances or technologies, the energy sources used by different appliances, and carbon emission factors for various energy sources.

Residential space cooling in urban and rural India primarily involves three types of household appliances: room ACs, fans, and air coolers [27], all of which operate on electricity. Therefore, the general formula of carbon intensity (i.e., carbon emissions per household) for Indian residential space cooling is as follows:

$$c = \sum_{x}^{OPTION} \sum_{y} Service\ volume \cdot Cooling\ intensity \cdot Energy\ efficiency \cdot Grid\ emission\ factor \quad (1)$$

where $x$ denotes the locale type (urban or rural) and $y$ denotes the type of appliance (room ACs, fans, or air coolers).

Given the distinct cooling principles employed by the three appliances, the specific meanings of the variables in the general formula Eq. (1) will be redefined for each appliance. Room ACs are appliances that rely on refrigerants in conjunction with other components of the air conditioning system to absorb and cool hot indoor air before the cool air is circulated back into the room. The service volume is defined as the amount of room floorspace supplied with room AC services per household. Currently, room ACs in Indian households predominantly include window and split ACs with a cooling capacity of 1-1.5 tons, designed to cool individual residential rooms of 15-20 square meters ($m^2$) [28]. Typically, Indian households have around three individual rooms, such as living rooms, bedrooms, or study rooms (excluding kitchens and toilets), which could potentially be equipped with air conditioning [29]. Based on the above information, this study assumed that one room AC can effectively serve 1/3 of the floorspace per household. The cooling intensity of room ACs is expressed as the annual cooling load per floorspace (unit: kWh/m$^2$·a). This parameter is primarily influenced by factors such as the cooling degree days of the area, the characteristics of the building envelope, and the total hours of annual cooling demand. Furthermore, the energy efficiency of room ACs is measured by the Indian seasonal energy efficient ratio (ISEER). The ISEER is defined as the ratio of the cooling seasonal total load to the cooling seasonal energy consumption and has been mandated for ACs in India since January 2018 [30]. In summary, the carbon intensity for room ACs in Indian residential space cooling $c_r$ is expressed as follows:

$$c_r = f_r \cdot l_r \cdot \lambda_r \cdot k \tag{2}$$

where $f_r$ denotes the room floorspace supplied with room AC services per household, $l_r$ denotes the annual cooling load per floorspace, $\lambda_r$ denotes ISEER, and $k$ denotes the grid emission factor.

For fans, which are nonrefrigerant appliances that rely on increasing airflow to make occupants feel cool (although fans alone do not actually cool the room air), the service volume is defined by the number of fans being used per household. The cooling intensity of fans is expressed as the annual usage hours of fans, which has little correlation with the floor space and temperature of rooms. The energy efficiency of fans is typically measured as the rate of air delivery achieved for each watt (W) of energy consumed. However, this study represented it differently by using its reciprocal, specifically, the power of fans at 210 m$^3$/min air delivery (unit: W) [31]. In summary, the carbon intensity for fans in Indian residential space cooling $c_f$ is expressed as follows:

$$c_f = u_f \cdot t_f \cdot \lambda_f \cdot k \tag{3}$$

where $u_f$ denotes the number of fans being used per household, $t_f$ denotes the annual usage hours of fans, and $\lambda_f$ denotes the power of fans at 210 m$^3$/min air delivery.

The principle of air coolers is similar to that of fans, where a water tank is added below a fan, causing the evaporation of water and cooling a small amount of air. The carbon intensity for air coolers in Indian residential space cooling is akin to that of fans and is expressed as follows:

$$c_a = u_a \cdot t_a \cdot \lambda_a \cdot k \tag{4}$$

where $u_a$ denotes the number of air coolers used per household, $t_a$ denotes the annual usage hours of air coolers, and $\lambda_a$ denotes the average rated power of air coolers commonly used in Indian households.

It should be noted that, due to the influence of India's tropical monsoon climate on residents' heat tolerance, the National Building Code suggests that the temperature threshold of 30°C is commonly regarded as the point at which occupants begin to feel uncomfortable [32]. Therefore, this study assumed that fans would be used when daily temperatures exceed 30°C. However, as temperatures rise further, people tend to increase their cooling capacity, transitioning from fans to more powerful solutions. Consequently, room ACs or air coolers are assumed to be used when daily temperatures exceed 35°C [33].

### *2.2. Top-down decomposition via the DSD method*

The carbon intensity associated with space cooling is influenced by three main factors: climate, population, and economy. India's tropical location, coupled with a monsoon climate, results in the entire territory being significantly influenced by high temperatures [34]. Consequently, the number of family members, denoted as the population per household, plays a pivotal role in influencing the selection of space cooling appliances and determining the spatial coverage and duration of cooling demand. The economic aspect is primarily characterized through income per capita, wherein the net state domestic product (NSDP) per capita serves as the official measure in India. A widely acknowledged perspective posits that the net domestic product offers a more holistic measure of social and economic well-being compared to the gross domestic product [35]. Finally, the emission factor is primarily employed to represent the average emission rate of carbon dioxide ($CO_2$) for household energy consumption. In summary, an identity equation for carbon intensity is derived from the aforementioned factors as follows:

$$c = \frac{C}{H} = \frac{P}{H} \cdot \frac{N}{P} \cdot \frac{E}{N} \cdot \frac{C}{E} = p \cdot n \cdot e \cdot k \tag{5}$$

where $c = \frac{C}{H}$ is the carbon intensity associated with space cooling, $p = \frac{P}{H}$ is the population per household, $n = \frac{N}{P}$ is the NSDP per capita, $e = \frac{E}{N}$ is the energy intensity (i.e., energy consumption per unit NSDP), and $k = \frac{C}{E}$ is the emission factor.

The carbon intensity of residential space cooling is intricately linked to the utilization of various cooling appliances. To investigate whether alterations in the adoption of residential space cooling appliances in India over the past two decades have substantially influenced changes in carbon intensity, this study introduced the DSD method, which was designed to effectively analyze the impact of structural changes [36]. Adhering to the fundamental principles of the DSD method [37], Eq. (6) is redefined as follows:

$$c = \sum_{i=1}^{3} p \cdot g \cdot n \cdot e \cdot w_i \cdot k_i \tag{6}$$

where $i$ represents the three main residential space cooling appliances, namely, room ACs, fans, and air coolers. $w_i$ is introduced as the share of $E_i$ in $E$, representing the energy consumption proportion of the three main appliances. For the equation $e_i = \frac{E_i}{N}$, it holds that $w_i = \frac{E_i}{E} = \frac{e_i}{e}$. $\sum_{i=1}^{3} w = 1$. $k_i$ is the emission factor of the energy consumption from appliance $i$. Since all three appliances consume electricity, it is obvious that $k_1 = k_2 = k_3$.

The total differential equations and the expanded matrix forms are represented in Eqs. (7-8):

$$\begin{cases} \mathrm{d}c = \sum_{i=1}^{3} \left( \frac{\partial c_i}{\partial p} \mathrm{d}p + \frac{\partial c_i}{\partial g} \mathrm{d}g + \frac{\partial c_i}{\partial n} \mathrm{d}n + \frac{\partial c_i}{\partial e} \mathrm{d}e + \frac{\partial c_i}{\partial w_i} \mathrm{d}w_i + \frac{\partial c_i}{\partial k_i} \mathrm{d}k_i \right) \\ \mathrm{d}w_i = \mathrm{d}F_i + \mathrm{d}F \\ \sum_{i=1}^{3} \mathrm{d}w_i = 0 \end{cases} \tag{7}$$

$$\begin{bmatrix} 1 & -\frac{\partial c_1}{\partial w_1} & -\frac{\partial c_2}{\partial w_2} & -\frac{\partial c_3}{\partial w_3} & 0 \\ 0 & 1 & 0 & 0 & -1 \\ 0 & 0 & 1 & 0 & -1 \\ 0 & 0 & 0 & 1 & -1 \\ 0 & 1 & 1 & 1 & 0 \end{bmatrix} \cdot \begin{bmatrix} dc \\ dw_1 \\ dw_2 \\ dw_3 \\ dF \end{bmatrix} = \begin{bmatrix} \sum \frac{\partial c_i}{\partial p} & \sum \frac{\partial c_i}{\partial g} & \sum \frac{\partial c_i}{\partial n} & \sum \frac{\partial c_i}{\partial e} & \frac{\partial c_1}{\partial k_1} & \frac{\partial c_2}{\partial k_2} & \frac{\partial c_3}{\partial k_3} & 0 & 0 & 0 \\ 0 & 0 & 0 & 0 & 0 & 0 & 0 & 1 & 0 & 0 \\ 0 & 0 & 0 & 0 & 0 & 0 & 0 & 0 & 1 & 0 \\ 0 & 0 & 0 & 0 & 0 & 0 & 0 & 0 & 0 & 1 \\ 0 & 0 & 0 & 0 & 0 & 0 & 0 & 0 & 0 & 0 \end{bmatrix} \cdot \begin{bmatrix} dp \\ dg \\ dn \\ de \\ dk_1 \\ dk_2 \\ dk_3 \\ dF_1 \\ dF_2 \\ dF_3 \end{bmatrix} \tag{8}$$

where $\mathrm{d}F_i$ and $\mathrm{d}F$ are introduced as artificial variables, breaking the changes in shares into two components, which are not linked to any level variables. $\mathrm{d}F$ is an endogenous slack variable, ensuring that the shares always sum to 1. To determine the impact of changes in the energy consumption share of an individual appliance, the corresponding shift variable $\mathrm{d}F_i$ is perturbed [37]. Moreover, the resulting change in $w_i$, $\mathrm{d}w_i$, is endogenously corrected through the adjustment of the slack variable $\mathrm{d}F$ to satisfy the unit sum constraint.

Given that the DSD method is based on Euler's method of numerical integration, this study partitioned the changes in exogenous variables into N segments, ensuring that $N$ is sufficiently large to make each segment small but not infinitesimal. This approach allows for a more approximate estimation of the cumulative impact of exogenous variables on endogenous variables. In accordance with the aforementioned principle, Eq. (8) is condensed into a generalized expression denoted as Eq. (9). Subsequently, the variations in each segment are represented by Eq. (10).

$$\boldsymbol{A} \cdot d\boldsymbol{y} = \boldsymbol{B} \cdot d\boldsymbol{z} \tag{9}$$

$$\begin{cases} \boldsymbol{D}^{(n)} = \left(\boldsymbol{A}^{(n-1)}\right)^{-1} \cdot \boldsymbol{B}^{(n-1)} \cdot diag(d\boldsymbol{z}) \\ d\boldsymbol{y}^{(n)} = \boldsymbol{D}^{(n)} \cdot \boldsymbol{j} \\ \boldsymbol{z}^{(n)} = \boldsymbol{z}^{(n-1)} + d\boldsymbol{z} \\ \boldsymbol{y}^{(n)} = \boldsymbol{y}^{(n-1)} + d\boldsymbol{y}^{(n)} \\ \boldsymbol{A}^{(n)} = f\left(\boldsymbol{z}^{(n)}, \boldsymbol{y}^{(n)}\right) \\ \boldsymbol{B}^{(n)} = g\left(\boldsymbol{z}^{(n)}, \boldsymbol{y}^{(n)}\right) \end{cases} \tag{10}$$

where $d\boldsymbol{z} = diag(d\boldsymbol{z}) \cdot \boldsymbol{j} = \frac{\Delta z}{N}$, $diag(d\boldsymbol{z})$ is the diagonal matrix composed of the differential element $d\mathbf{y}$, and $\mathbf{j}$ is a vector consisting of ones. For $n = 1,2,\cdots,N$, this study set N=16000 to ensure adequate precision, as stipulated in the original research article on the DSD method [37].

The desired decomposition is achieved through iterative computations, with contributions from each change being summed:

$$\left\{ \mathbf{D} = \sum_{n=1}^{N} \mathbf{D}^{(n)} \right. \tag{11}$$

In summary, the decomposition of changes in carbon intensity for residential space cooling adheres to the method described above, which yields the following:

$$\begin{cases} \Delta c|_{0\to T} = \Delta p_{\mathrm{DSD}} + \Delta g_{\mathrm{DSD}} + \Delta n_{\mathrm{DSD}} + \Delta e_{\mathrm{DSD}} + \Delta k_{\mathrm{DSD}} + \Delta w_{\mathrm{DSD}} \\ \Delta k_{\mathrm{DSD}} = \sum_{i=1}^{3} \Delta k_i \\ \Delta w_{\mathrm{DSD}} = \sum_{i=1}^{3} \Delta w_i \end{cases} \tag{12}$$

## *2.3. Variable identification and data collection*

Table 1 details the variables presented in Sections 2.1 and 2.2, along with their definitions, units, and sources for data collection.

**Table 1.** Variable interpretation: definitions, units, and data collection.

| Variable | Definition | Unit | Data collection | |
|---|---|---|---|---|
| | | | Data source categorization | Data references and derivation |
| $f_r$ | Room floorspace supplied with room AC services per household | $m^2$ | Official statistics | • Residential building floorspace data: IEA [38]<br>• Ownership data for room ACs: MoEFCC India [26]<br>• Assumption that one room AC can effectively serve 1/3 of the floorspace per household: MoHFW India [7] |
| $l_r$ | Annual cooling load per floorspace | $kWh/m^2 \cdot a$ | Research papers | • Urban building cooling load data: Arul Babu et al. [39]<br>• Rural building cooling load data: Celestial Earth [40] |
| $t_a$ | Annual usage hours of air coolers | Hours (h)/year | Research papers | • Air cooler usage data: Walia et al. [41]<br>• Assumption on air cooler usage when daily temperature exceeds 35 °C: Morris et al. [33] |
| $t_f$ | Annual usage hours of fans | h/year | Research papers | • Fan usage data: Walia et al. [41]<br>• Assumption on fan usage when daily temperature exceeds 30 °C: Morris et al. [33] |
| $u_a$ | Number of air coolers being used per household | Million units | Official statistics & Institutional report | • Air cooler stock data: MoEFCC India [26]<br>• Ownership data for air coolers by state: LBNL [42] |
| $u_f$ | Number of fans being used per household | Million units | Official statistics & Institutional report | • Fan stock data: MoEFCC India [26]<br>• Ownership data for fans by state: Prayas [43] |
| $\lambda_a$ | Average rated power of air coolers commonly used in Indian households | W | Institutional report | • Power data of air coolers: AEEE [44] |
| $\lambda_f$ | Power of fans at 210 $m^3$/min air delivery | W | Institutional report | • Power of fans: Prayas [45] |
| $\lambda_r$ | ISEER | - | Institutional report | • ISEER of room ACs: TERI [14] |
| $P$ | Population size | Million persons | Official statistics | • Population size data: Indiastat [46] |
| $H$ | Amounts of households | Million households | Official statistics | • Households data: MoHFW India [7] |
| $N$ | NSDP | Indian Rupee (INR) | Official statistics | • NSDP data: Indiastat [47] |
| $c$ | Carbon intensity (carbon emissions per household) associated with residential space cooling | Kilograms of carbon dioxide ($kgCO_2$) per household | Calculated by authors | $c = c_r + c_f + c_a$ |
| $C$ | Carbon emissions from residential space cooling | Mega-tons of carbon dioxide ($MtCO_2$) | Calculated by authors | $C = c \times H$ |
| $k$ | Emission factors | $tCO_2/kWh$ | Official statistics | • Grid carbon emission factor data: IEA [48] |
| $E$ | Energy consumption associated with residential space cooling | Gigawatt hours (GWh) | Calculated by authors | $E = \frac{C}{k}$ |
| $p$ | Population per household | Persons per household | Calculated by authors | $p = \frac{P}{H}$ |
| $n$ | NSDP per capita | INR/person | Calculated by authors | $n = \frac{N}{P}$ |
| $e$ | Energy intensity (energy consumption per unit NSDP) | kWh/INR | Calculated by authors | $e = \frac{E}{N}$ |
| $w$ | Energy structure of the three main appliances (room ACs, fans, and air coolers) | % | Calculated by authors | $w_i = \frac{e_i}{e}$ |

## 3. Results

Building on the materials and methods outlined in Section 3, this section focuses on two key results of the study: the historical evolution of carbon intensity in residential space cooling and the decomposition of carbon intensity changes in residential space cooling in India.

### *3.1. Historical evolution of carbon intensity for residential space cooling*

This study utilized a bottom-up approach to investigate the historical carbon intensity growth of space cooling in Indian households from 2000 to 2022, exploring the respective contributions of three main types of space cooling appliances—room ACs, fans, and air coolers.

As illustrated in Fig. 2 a, the carbon intensity for space cooling in Indian households increased by 292.4% from 2000 to 2022, with an average annual growth rate of 6.4%, reaching an average carbon intensity of 513.8 kg$CO_2$ per household in 2022. When examining the individual performance of each state, Sikkim emerged with the highest growth rate, increasing by 530.0% from 2000 to 2022, with an average annual growth rate of 8.7%. Despite this remarkable surge, Sikkim maintained the lowest carbon intensity for residential space cooling in the country, recording a mere 43.3 kg$CO_2$ per household in 2022. The important factor contributing to Sikkim's distinctive trends in residential space cooling is its geographical location in the Himalayas, affording it a temperate climate during the summer months, characterized by average temperatures ranging from 15 to 25°C. Significantly, Sikkim stands out as one of India's most affluent states, boasting the lowest poverty levels nationwide [49]. In 2022, the NSDP per capita of urban households in Sikkim reached 384,891 rupees, the highest in the country. The intersection of prosperity and urbanization has notably heightened the demand for residential space cooling in this region. Furthermore, the 21st century phenomenon of global climate warming has initiated the melting of glaciers in the Himalayas, resulting in occasional extreme heat waves in Sikkim. This climatic shift further accelerated the surge in cooling demand within the state.

In contrast, Chandigarh exhibited the lowest growth rate, at only 113.0% from 2000 to 2022, with an average annual growth rate of 3.5%. Strikingly, it boasted the highest carbon intensity in the country, reaching 1826.5 $kgCO_2$ per household in 2022. This can be attributed to Chandigarh having the highest per-household availability of ACs in India, earning the reputation of being a renowned AC city. According to the fifth National Family Health Survey in 2021 [7], 77.9% of urban households in Chandigarh have room ACs, surpassing figures for Delhi (74.3%), Punjab (70.2%), and Haryana (61.8%), all of which are located in the northwestern region of India. This region consistently experiences abnormally hot days from March to June, occasionally extending into July. Furthermore, Chandigarh boasted the highest urbanization rate among Indian states and union territories, reaching 99.8% in 2022. The saturation of urbanization and a high share of ACs have consequently led to a plateau in residential space cooling demand, exhibiting limited upward momentum.

In general, due to the hot climate and frequent heatwaves across India, except for the relatively cooler summer climate in the Himalayan North region and the high-altitude areas on the Deccan Plateau (particularly Bangalore in Karnataka), the average summer temperature in other parts of the country soars as high as 35 °C, necessitating cooling demand from March to November. Consequently, the carbon intensity for residential space cooling in each state was influenced mainly by the urbanization level in 2022, as urban households have greater access to cooling appliances. The analysis of the growth rates from 2000 to 2022 revealed that higher rates were primarily observed in northern India and the southern Deccan Plateau. These regions, which are traditionally known for their temperate summer weather, have experienced an increase in extreme heatwaves due to climate change in the 21st century. This climatic shift has spurred robust growth in the space cooling demands of local households.

The increase in carbon intensity for space cooling in Indian households was examined from 2000 to 2011 and 2011 to 2022, with a detailed analysis conducted at both the urban and rural levels. The analysis included a decomposition of the contributions from three main space cooling appliances, as shown in Fig. 2 b.

From 2000 to 2011, there was a substantial increase in carbon intensity for space cooling in Indian urban households, surging from 285.1 $kgCO_2$ per household to 733.1 $kgCO_2$ per household, marking a 157.1% increase. Analyzing the contributions of the three main cooling appliances, 97.7% of this increase was attributed to fans, with a carbon intensity increase of 437.6 $kgCO_2$ per household from fans. During this period, there was an average increase of 1.6 fans per Indian urban household. In the subsequent period from 2011 to 2022, minimal growth in carbon intensity was observed, reaching 744.7 $kgCO_2$ per household in 2022. Room ACs emerged as the most positive contributor and were responsible for a carbon intensity increase of 43.2 $kgCO_2$ per household. The increase in carbon intensity for air coolers amounted to 17.9 $kgCO_2$ per household. Conversely, fans exhibited a negative contribution, showing a decline in carbon intensity from 573.9 to 524.5 $kgCO_2$ per household during the 2011 to 2022 period. This indicates that space cooling from fans reached a saturation point in urban Indian households, despite the fact that fans still contribute 70.4% of the carbon intensity for space cooling in urban households, with ubiquitous utilization across all income strata, including homes equipped with room ACs [50]. While room ACs currently have a low penetration rate (approximately 12%) in urban Indian households, the trajectory toward refrigerant-based appliances, particularly room ACs with enhanced cooling effects, is expected to grow significantly, potentially replacing fans in the coming decades. Furthermore, the growth of air coolers, which are nonrefrigerant-based appliances that cool the atmosphere through water evaporation, is limited because their effectiveness is constrained by the humidity of the ambient air, and these coolers exhibit optimal performance in hot and dry climates and in composite climates.

In 2000, nearly no rural households in India owned room ACs, and the carbon intensity for space cooling households was 68.5 $kgCO_2$ per household. By 2011, it had surged to 290.0 $kgCO_2$ per household, reflecting a growth rate of 323.2%, with 91.2% of the increase still attributed to fans. Subsequently, from 2011 to 2022, the carbon intensity for space cooling in rural Indian households experienced an increase of 23.4%, reaching 358.1 $kgCO_2$ per household in 2022. Fans continued to dominate, contributing 46.1% to the overall increase in carbon intensity, followed by air coolers at 38.6%. Despite a small number of affluent rural households gradually adopting room ACs to cope with extreme heat waves (with a penetration rate of approximately 1.2%), the carbon intensity from room ACs remained low at 20.2 $kgCO_2$ per household in 2022, constituting less than 10% of the 294.6 $kgCO_2$ per household from fans. Although India has achieved near-complete electricity access [51], the transformation of space cooling demand in rural households is expected to take at least 20 years, requiring initial impetus from urban areas. Insights into the changes in room AC ownership in China's urban and rural areas from 2000 to 2020 can provide a reference for this transition.

Overall, this microlevel investigation into the utilization of residential space cooling appliances during building operations, which examined the historical evolution of carbon intensity for space cooling from 2000 to 2022 in urban and rural households across the Indian state, addresses Question 1 outlined in Section 1.

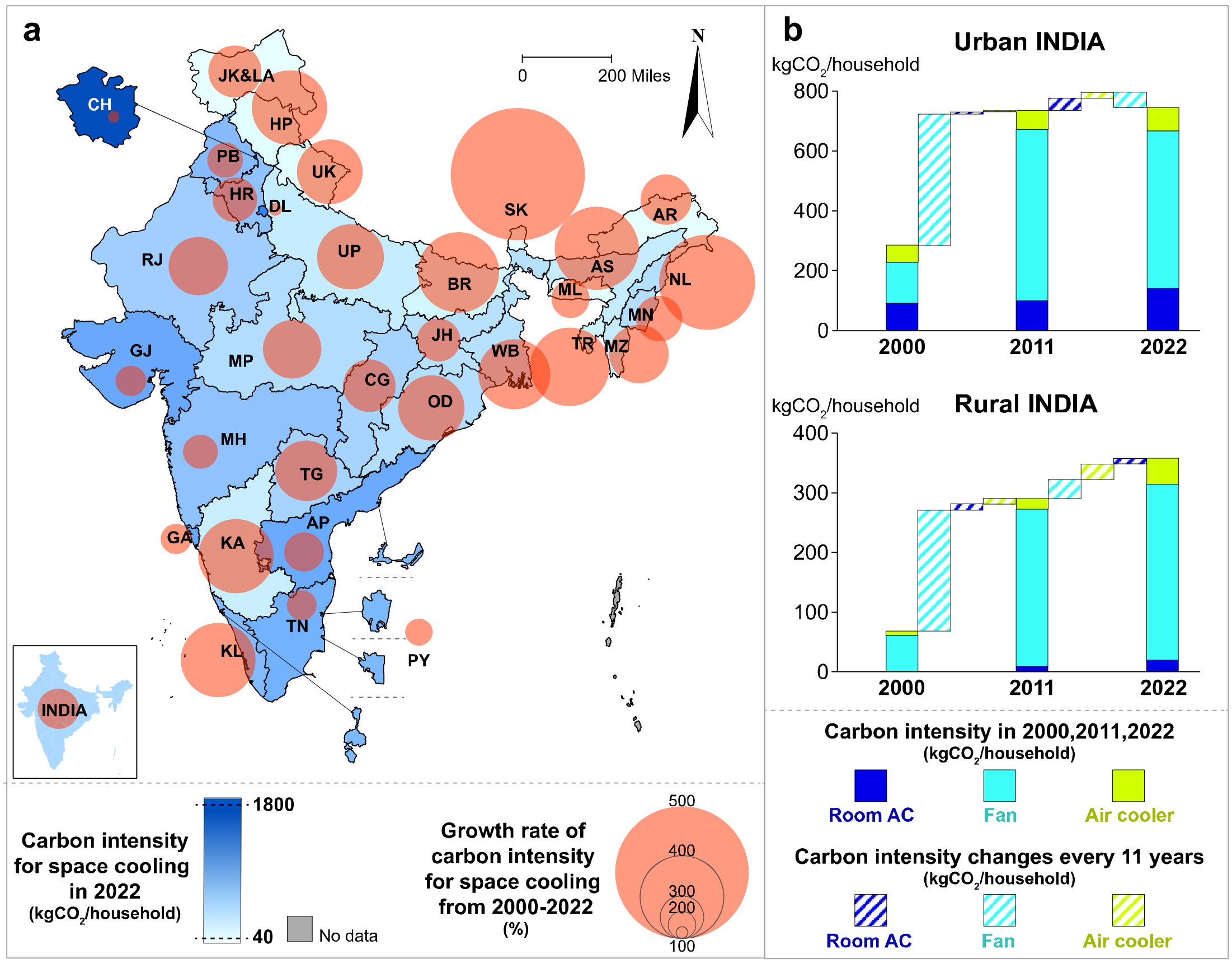

**Fig. 2.** (a) Carbon intensity for space cooling in Indian households from 2000 to 2022 across different states; (b) Carbon intensity of the three main space cooling appliances in both urban and rural households. Note: the full names of the Indian states and union territories are detailed in Appendix B.

### *3.2. Decomposition of carbon intensity changes in residential space cooling*

This study employed the DSD method to decompose the changes in carbon intensity for residential space cooling from 2000 to 2022 considering five factors: NSDP per capita, emission factors, population per household, energy intensity, and cooling appliance structure.

Fig. 3 a illustrates the decomposition of the carbon intensity change for residential space cooling during 2000-2011 and 2011-2022. It provides a detailed breakdown of the results at the national level, distinguishing between urban and rural households. During the 2000-2011 period, carbon intensity experienced a significant surge, particularly in Indian rural households. During the subsequent period from 2011 to 2022, the growth rate notably slowed, primarily due to the saturation of demand for fans, while the buying power for room ACs and air coolers remained sluggish, as explained in Section 3.1.

Further analysis of the driving factors influencing changes in carbon intensity revealed that the NSDP per capita consistently made the most positive contribution from 2000 to 2022, with a contribution rate of 196.5% for all Indian households, 262.2% for urban households, and 163.1% for rural households. The exception is 2021, where Indian national income experienced a sharp decline due to COVID-19, leading to a rare occurrence of NSDP per capita having a negative contribution to the change in carbon intensity. This observation underscores that the enhancement of cooling demand in residential buildings is substantially driven by the improvement in family economic status. The emission factors were the most significant negative contributors over the past 22 years, with a negative contribution rate of -41.5% for all Indian households, -55.1% for urban households, and -36.7% for rural households. This underscores the persistent efforts in the clean energy transition within India's power sector. The negative impact of population per household on carbon intensity also persisted and was particularly pronounced in urban households, where the contribution rate was -40.5%. This was intricately linked to the ongoing reduction in family size observed in India. Moreover, energy intensity, denoting energy consumption per unit of NSDP, initially contributed positively to the growth of carbon intensity during the 2000-2011 period but transitioned into a negative contribution from 2011 to 2022. This shift was primarily attributed to the gradual deceleration in energy consumption growth during the latter period, which decreased from an annual average growth rate of 16.2% to a more moderate 5.5%. In addition, household income maintained a consistent annual average growth rate of 9.9%, resulting in a decline in energy intensity. Finally, the cooling appliance structure, which represents the energy consumption proportion of the three main appliances, had a minimal effect on the change in carbon intensity. This was attributed to the continuous dominance of fans in the space cooling supply. The energy consumption of room ACs and air coolers accounted for less than a quarter of the total for the past two decades, indicating that the overall cooling appliance structure did not undergo significant changes.

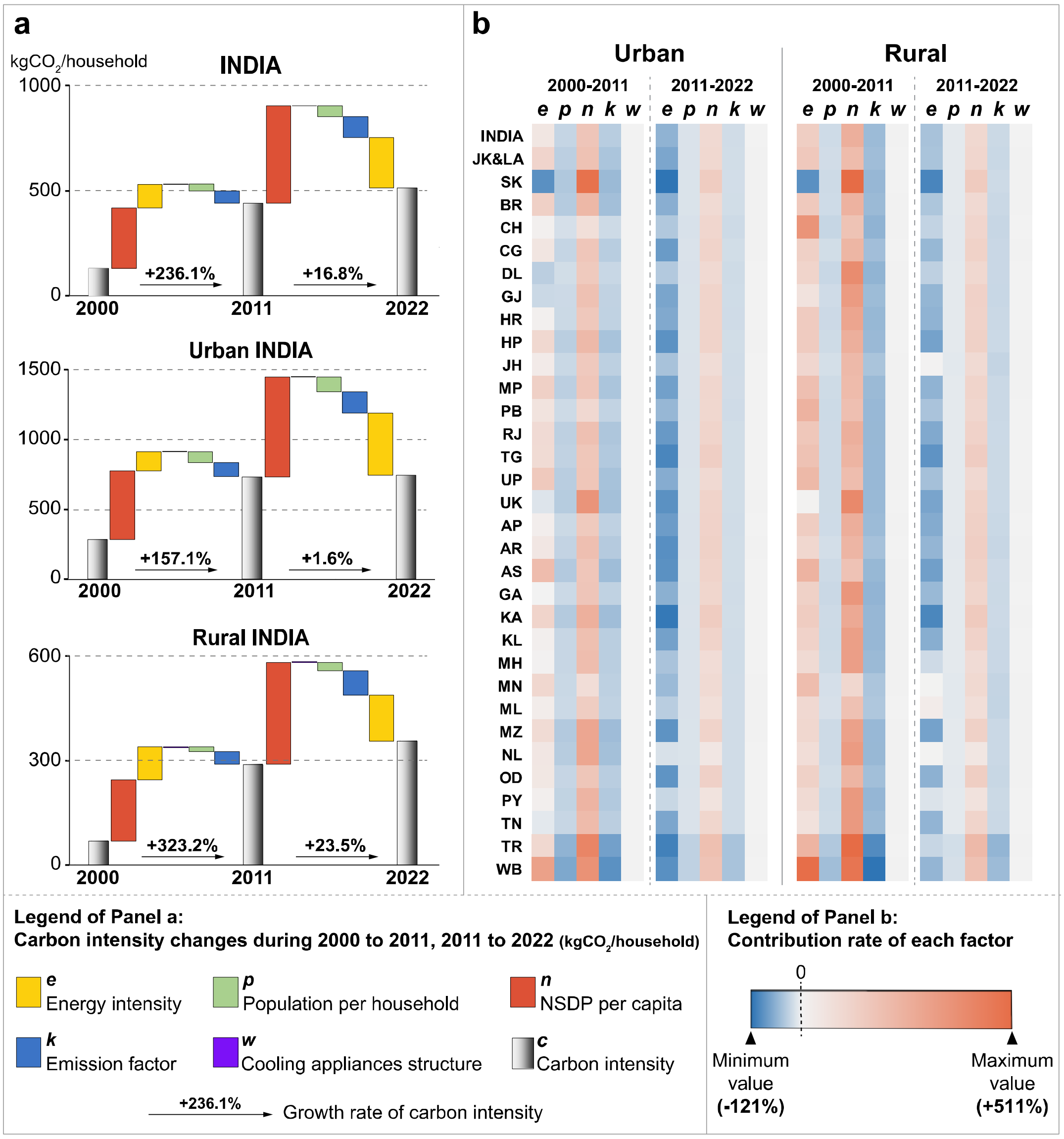

**Fig. 3.** Decomposition of carbon intensity changes for space cooling in Indian households from 2000 to 2011 and from 2011 to 2022 at the (a) national level and (b) state level.

Fig. 3 b further illustrates the decomposition of carbon intensity changes in rural and urban households across various Indian states. The results of the decomposition for most states were consistent with those at the national level in 2000-2011 and 2011-2022. The energy intensity of urban households in states such as Sikkim, Delhi, Gujarat, Uttarakhand, and Tamil Nadu showed a negative contribution to the growth of carbon intensity from 2000 to 2011. This suggests that the energy intensity of urban households in these states started to exhibit a downward trend as early as 10 years ago. Furthermore, rural households in Sikkim exhibited similar characteristics. The state consistently maintained relatively low cooling energy consumption and is recognized as the best-performing small state in terms of the economy. There was a remarkable 1317.1% increase in the rural NSDP from 2000 to 2022 (at constant prices), doubling nationwide growth in India.

Overall, the macrolevel exploration delved into societal, economic, climatic, and technological factors influencing changes in carbon intensity for India's residential space cooling, partially addressing Question 2 outlined in Section 1.

## 4. Discussion

This section focuses on several key implications of the study, including the decarbonization performance of residential space cooling across Indian states, the decarbonization strategy trajectories of residential space cooling appliances, and the building efficiency pathways for decarbonizing residential space cooling in India.

### *4.1. Decarbonization performance in residential space cooling across Indian states*

The decarbonization of building operations involves efforts to mitigate carbon emissions while meeting regular energy consumption requirements to satisfy the housing demands of occupants. This implies that the estimated operational carbon emissions should align with the observed carbon emissions plus the estimated decarbonization. The concept similarly applies to decarbonization intensity. Consistent with the literature [24], this study defined decarbonization intensity ($\Delta Dc|_{0\to T}$) for space cooling in Indian households during a given period as the sum of the negative contributions from various factors identified in the decomposition of carbon intensity changes. It can be expressed in Eq. (13):

$$\Delta Dc|_{0\to T} = -\sum\left(\Delta c(X_j)|_{0\to T}\right) \tag{13}$$

Where $\Delta c\left(X_j\right) \in \left(\Delta e_{DSD}, \Delta p_{DSD}, \Delta n_{DSD}, \Delta k_{DSD}, \Delta w_{DSD}\right), \Delta c(X_j)|_{0\to T} < 0$

Fig. 4 a illustrates the cumulative decarbonization intensity for space cooling of urban and rural households across Indian states from 2001 to 2022. Overall, the cumulative decarbonization intensity for space cooling in Indian urban households reached 1386.1 $kgCO_2$ per household over these 22 years, while that for rural households was 368.8 $kgCO_2$ per household. The pie chart above the map indicates that the cumulative decarbonization intensity for residential space cooling from 2012 to 2022 constituted 72.2%, highlighting strong decarbonization efforts on household cooling consumption in India over the past decade. The disparity in decarbonization intensity for space cooling in urban and rural households, as well as the observed proportions in the two stages across states, aligns with the features at the national level. Goa had the highest decarbonization intensity for space cooling in both urban and rural households. The cumulative decarbonization intensity in urban households from 2001 to 2022 was 3172.4 $kgCO_2$ per household, which was 2.3 times the national average. Similarly, the cumulative decarbonization intensity in rural households was 1698.1 $kgCO_2$ per household, surpassing the national average by 4.6 times. States with higher decarbonization intensity were concentrated in areas with greater ownership of residential space cooling appliances, such as Goa, Chandigarh, Delhi, Andhra Pradesh, Gujarat, Tamil Nadu, Puducherry, and Kerala. The cumulative decarbonization intensity in these regions exceeded 1800 $kgCO_2$ per household from 2001 to 2022, and the cumulative decarbonization intensity per rural household was above 1000 $kgCO_2$ per household. On the other hand, states with lower decarbonization intensities were concentrated in the Indian Himalayan region, which is known for its mild summer climate. These states include Jammu & Kashmir & Ladakh, Himachal Pradesh, Uttarakhand, Meghalaya, Sikkim, and Arunachal Pradesh. The cumulative decarbonization intensity in urban areas in these regions was less than 300 $kgCO_2$ per household from 2001 to 2022, while the cumulative decarbonization intensity in rural areas was generally less than 100 $kgCO_2$ per household.

Fig. 4 b illustrates the cumulative decarbonization per unit of NSDP for space cooling in urban and rural households across Indian states from 2001 to 2022. Nationally, the decarbonization per unit of NSDP for space cooling in Indian urban households over these 22 years amounted to 379.7 $kgCO_2$ per lakh rupee, and for rural households, it was 218.4 $kgCO_2$ per lakh rupee. In addition, the cumulative decarbonization per unit of NSDP in both Indian urban and rural households during 2001-2011 and 2012-2022 remained essentially the same. However, in states with accelerated economic growth earlier, such as Sikkim, Uttarakhand,

Nagaland, Gujarat, Puducherry, and Tamil Nadu, the decarbonization per unit of NSDP in urban and rural households was notably greater in the former period than in the latter period. When comparing the decarbonization per unit of NSDP for residential space cooling among states, Bihar achieved the highest value in urban households from 2001 to 2022, reaching 823.8 $kgCO_2$ per lakh rupee. This figure was 2.2 times greater than the national average. Similarly, Andhra Pradesh recorded the highest value in rural households, totaling 546.2 $kgCO_2$ per lakh rupee, which was 2.5 times the national average. In general, states with greater decarbonization per unit of NSDP for space cooling were found in regions with higher annual average temperatures and lower household incomes, such as Bihar, Andhra Pradesh, Rajasthan, and Madhya Pradesh. The cumulative decarbonization per unit of NSDP in urban households in these states from 2001 to 2022 exceeded 600 $kgCO_2$ per lakh rupee, while in rural households, it surpassed 400 $kgCO_2$ per lakh rupee. Conversely, states with lower decarbonization per unit of NSDP were predominantly situated in the Indian-Himalayan region. The decarbonization per unit of NSDP in urban and rural households from 2001 to 2022 in these areas was less than 100 $kgCO_2$ per lakh rupee.

The analysis of decarbonization intensity and decarbonization per unit of NSDP across states underscores that decarbonization efforts should extend beyond an exclusive focus on household income or climate. In addition to prioritizing areas characterized by a higher space cooling supply, indicative of high household income and substantial ownership of cooling appliances, equal attention should be given to areas with pronounced cooling demand that is yet unmet, characterized by low household income but a hot climate.

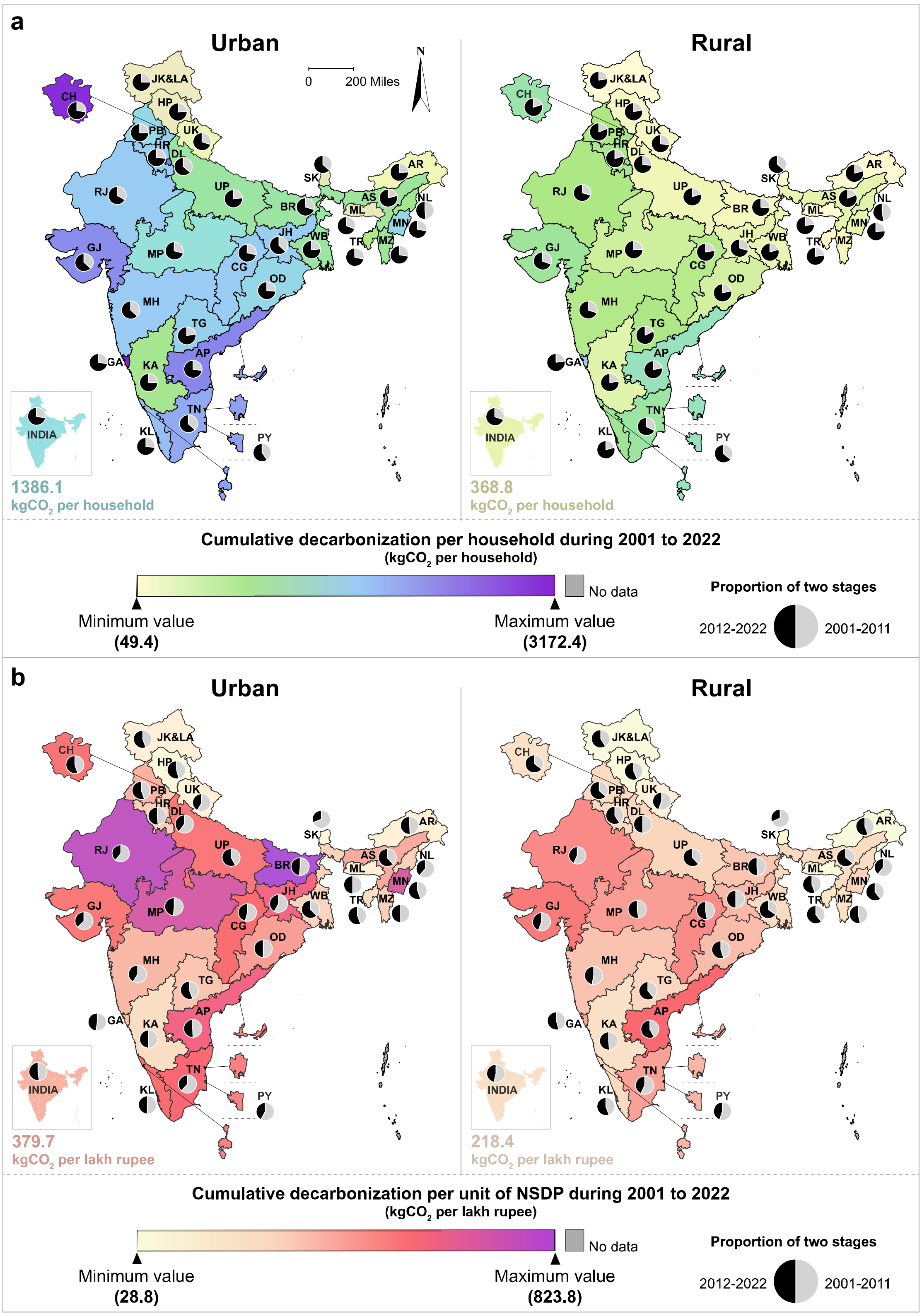

**Fig. 4.** (a) Decarbonization intensity and (b) decarbonization per unit of NSDP for space cooling in India's urban and rural households from 2001 to 2022.

As previously discussed, the decarbonization performance of states is intricately linked to their climate and income levels. To mitigate biases in assessing decarbonization capabilities arising from variations in geographical, economic, and other characteristics, this study introduced a relative metric: decarbonization efficiency. This metric was employed for a more objective evaluation of each state's historical efforts and future potential for decarbonization. The decarbonization efficiency ($\Delta d|_{0\to T}$) for a given period was calculated by dividing the total decarbonization amount ($\Delta DC|_{0\to T}$) in that period by the corresponding carbon emissions, which can be expressed as Eqs. (14-15):

$$\Delta DC|_{0\to T} = H|_{0\to T} \times (\Delta Dc|_{0\to T}) \tag{14}$$

$$\Delta d|_{0\to T} = \frac{\Delta DC|_{0\to T}}{C|_{0\to T}} \tag{15}$$

From 2001 to 2022, the total cumulative decarbonization for space cooling in Indian households amounted to 206.2 megatons of carbon dioxide ($MtCO_2$), achieving a decarbonization efficiency of 8.5%. The cumulative decarbonization efficiency is defined as the ratio of the cumulative decarbonization to the corresponding carbon emissions calculated annually starting in 2001. India's cumulative decarbonization efficiency has experienced rapid growth since 2006, reaching a peak of 11.2% in 2012. Following a slight decrease, it remained stable, fluctuating between 8 and 10% over the next 10 years.

Fig. 5 a depicts the evolution of cumulative decarbonization and the corresponding decarbonization efficiency for each state over the past 22 years. Due to substantial variations in population, climate, and economy among states, the magnitude of cumulative decarbonization exhibited significant differences. Consequently, Fig. 5 a utilizes four subfigures to illustrate the decarbonization development of states with varying magnitudes of cumulative decarbonization. This indicates that Maharashtra achieved the highest cumulative decarbonization from 2001 to 2022, totaling 28.9 $MtCO_2$, while Sikkim had the smallest amount, with only 0.014 $MtCO_2$. Since 2016, the cumulative decarbonization efficiency of most states has consistently stabilized at approximately 10%. Goa had the highest cumulative decarbonization efficiency, reaching 14.9% from 2001 to 2022, with the peak occurring in 2015 at 20.2%. Goa has taken the lead in the State Energy and Climate Index assessment [52], a tool designed to track the climate and energy-related efforts of states and union territories. Concurrently, the Goa government has declared its commitment to achieving net-zero emissions by 2050, aspiring to be a pioneer in India across all aspects of energy conservation and carbon reduction [53].

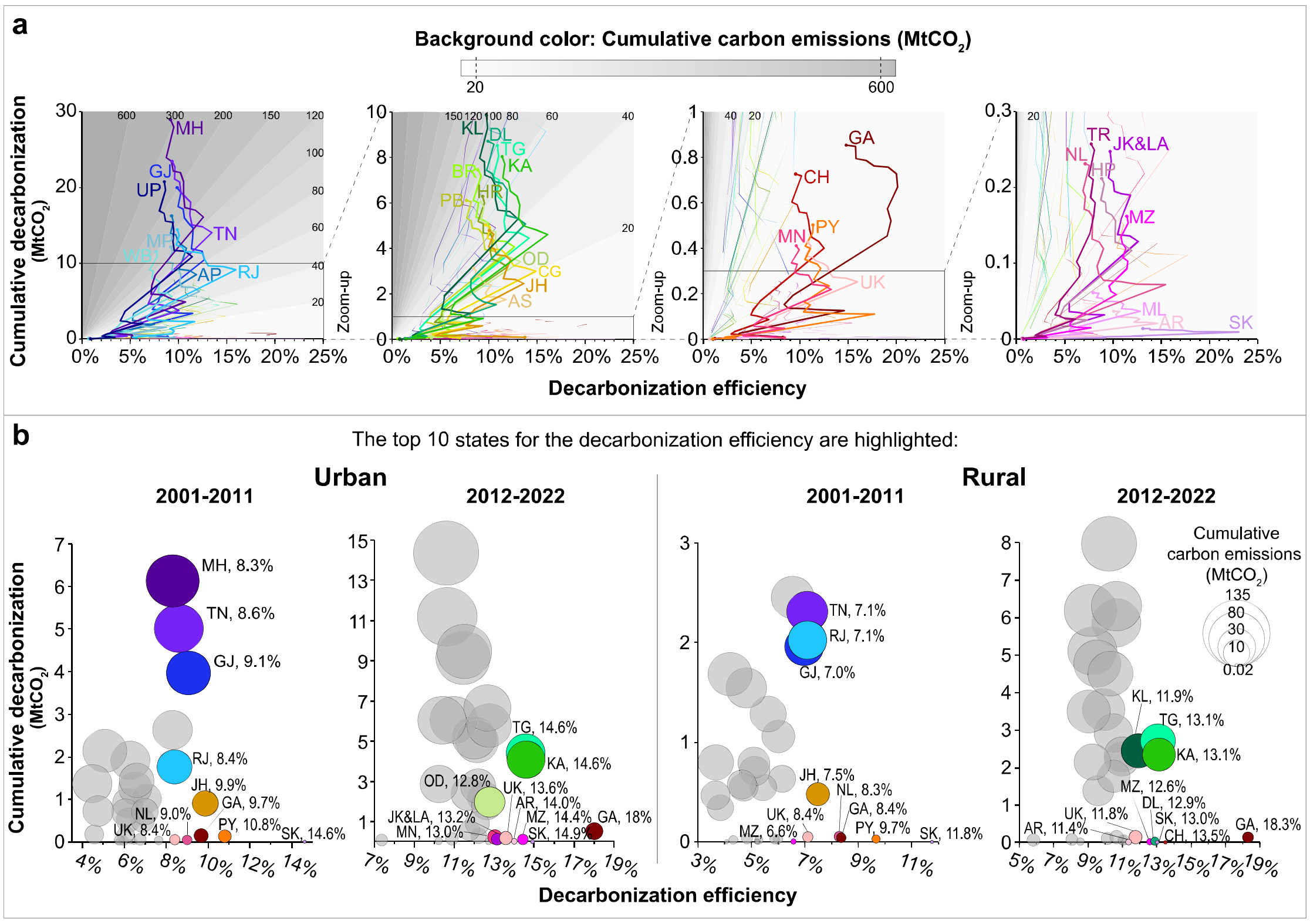

**Fig. 5.** (a) Cumulative decarbonization performance in residential space cooling across states; (b) Detailed representation of the space cooling decarbonization performance in urban and rural households from 2001-2011 and 2012-2022.

Fig. 5 b provides a detailed representation of the space cooling decarbonization performance of both urban and rural households across each state during the two periods spanning from 2001-2011 and 2012-2022. From 2001 to 2011, the decarbonization efficiency of urban Indian households was 5.9%, while that of rural households was 4.7%. When comparing states, the top ten rankings of decarbonization efficiency in urban households were primarily concentrated in a range of 8% to 10%, whereas in rural households, the top ten rankings were mainly in the 7% to 9% range. Tamil Nadu and Gujarat emerged as states with both higher decarbonization and decarbonization efficiency in both rural and urban households. In the subsequent period from 2012 to 2022, the decarbonization efficiency in Indian urban households increased to 11.1%, while in rural households, it reached 6.8% in both rural and urban households. The top ten decarbonization efficiency rankings in urban households across states significantly increased, ranging from 13% to 15%, whereas in rural households, they ranged from 11% to 14%. Notably, provinces with greater cumulative decarbonization, such as Maharashtra, Tamil Nadu, and Uttar Pradesh, did not exhibit advantages in terms of decarbonization efficiency. Instead, higher decarbonization efficiency was more concentrated in states where carbon emissions and decarbonization ranked in the middle or lower levels, such as Telangana, Karnataka, and Goa.

Overall, the aforementioned examination of decarbonization performance in residential space cooling across Indian states further addresses Question 2 outlined in Section 1.

### *4.2. Decarbonization strategy trajectories of residential space cooling appliances in India*

In response to the formidable challenges posed by the escalating heat crisis affecting India's society and economy, the Ministry of Environment, Forest and Climate Change launched the ICAP in March 2019, marking the world's pioneering initiative in formulating a comprehensive national cooling strategy. The ICAP represents a long-term vision with the primary objective of reducing the cooling power demand across various sectors in India by 20-25% by the year 2038. This ambitious goal is to be realized through the optimization of cooling systems, the enhancement of energy efficiency, the transition to sustainable refrigerants, and the incorporation of advanced technologies [26]. Specifically, concerning residential space cooling, the ICAP has functioned as a continuation and reinforcement of the Indian government's efforts to enhance the energy efficiency of space cooling in buildings over the past two decades. It also serves as a preemptive measure against the anticipated surge in household cooling demand in India over the next two decades.

As illustrated in Fig. 6, this study summarized the decarbonization strategy trajectories for residential space cooling appliances in India, covering both past trends and future projections. From a review perspective, it evaluated the decarbonization efforts in residential space cooling over the past two decades, focusing on improvements in the energy efficiency of key appliances—fans, room ACs, and air coolers—as depicted in Fig. 6 a. Looking ahead, the study also predicted the carbon emission trajectories of these three appliances through 2030, based on the policy goals outlined in the ICAP, as shown in Fig. 6 b.

Since 2000, room ACs have been central to the Indian government's policies and strategies aimed at enhancing the energy efficiency of residential space cooling, and were among the first appliances included in India's Standards and Labelling (S&L) scheme [54]. By 2009, fixed-speed room ACs have become subject to mandatory S&L regulations. Over time, advancements in air conditioning technology have prompted multiple revisions to the star rating labels for the two primary types of room ACs, split ACs and window ACs, progressively elevating the standards [55]. In 2018, India transitioned from its original energy efficiency standards for air conditioning systems to a new standard, ISEER [30]. Simultaneously, the S&L scheme for variable-speed room ACs was made mandatory. While there has been initial success in achieving comprehensive standardization of energy efficiency for residential room ACs in India, as of 2022, the number of room ACs remained below 20 units per 100 households, with an average electricity consumption of 88.7 kWh per household annually. Due to historically modest demand for room ACs in Indian households over several decades, the improvement in energy efficiency standards has resulted in a cumulative reduction of approximately 120 Mt$CO_2$ in carbon emissions over the past 22 years. However, there is an anticipation of a significant surge in the number of room ACs in Indian households, with a projected increase of 2-6 times in the next 10-20 years. In response, the ICAP proposed a steady annual growth of 3% in the efficiency levels (i.e., ISEER) of room ACs moving forward [26].

Fans continue to be the predominant residential space cooling appliance in India, largely due to the prohibitive costs of acquiring and operating room ACs. As of 2022, the number of household fans in India reached an impressive 180 units per 100 households, with an average electricity consumption of 493.9 kWh per household annually, 5.6 times greater than that of room ACs. Nevertheless, the cumulative reduction in carbon emissions resulting from enhanced energy efficiency standards and the modest increase in the market share of energy-efficient fans over the past 22 years was approximately 300 Mt$CO_2$, which was only 2.5 times that of room ACs. The infrequency of government control and the enhancement of energy efficiency standards for fans, coupled with limited consumer awareness regarding the cost-saving benefits of energy-efficient fans, contributed to this disparity. Ceiling fans were initially included in the voluntary scope of the S&L scheme in 2010. Their star ratings of energy efficiency standards were revised once in 2019 under IS 374:2019 before eventually moving to the mandatory scope only in 2022 [56]. Despite the ban on the sale and production of unlabeled fans in India beginning in 2023, a survey conducted by Prayas [45] revealed that unlabeled fans are still widely available in stores. Consumer preferences often prioritize price and aesthetics over star ratings. A study by the Council on Energy, Environment, and Water indicated that enthusiasm for star-rated fans has been low, with only three percent of households surveyed owning one [57]. Recognizing the need for a substantial shift, the Energy Efficiency Services Limited in India planned to purchase and distribute 10 million efficient Brushless Direct Current fans in 2023, representing one-fourth of the market sales. This initiative aims to boost consumer acceptance of energy-efficient fans at an affordable price and stimulate significant replacement demand for fans across India. Prior to this, the company achieved success in driving down the prices of energy-efficient light-emitting diode (LED) bulbs through bulk procurement, contributing to the achievement of the Unnat Jyoti by Affordable LEDs for All scheme launched in 2014 [58]. The ICAP emphasizes the complete replacement of conventional fans (approximately 70W) with energy-efficient ceiling fans (approximately 50W) over the next decade [26]. This initiative is considered a crucial step in India's ambitious commitment to achieving net-zero emissions by 2070.

Air coolers have experienced a noticeable surge in sales in recent years, particularly among low- and middle-income households. This growth is driven by escalating heatwaves, improved power supply in villages, and the significant 80%-85% cost advantage over room ACs. Despite this surge in demand, the energy performance standards for air coolers outlined in IS 3315 have remained unchanged since 1994. Additionally, air coolers are not currently covered in the S&L scheme [44]. Currently, the air cooler market remains largely under the control of unorganized local manufacturers, offering cheap yet nonstandard and highly inefficient models. Nevertheless, the introduction of the Goods and Services Tax in India has facilitated the entry of branded air coolers into the market, narrowing the price gap and promoting standardization. This development is crucial for the potential future inclusion of air coolers in the S&L scheme. The ICAP calls for the introduction of Minimum Energy Performance Standards (MEPS) to reduce the typical energy consumption of residential air coolers (approximately 200W) by 10%- 20% over the next decade [26].

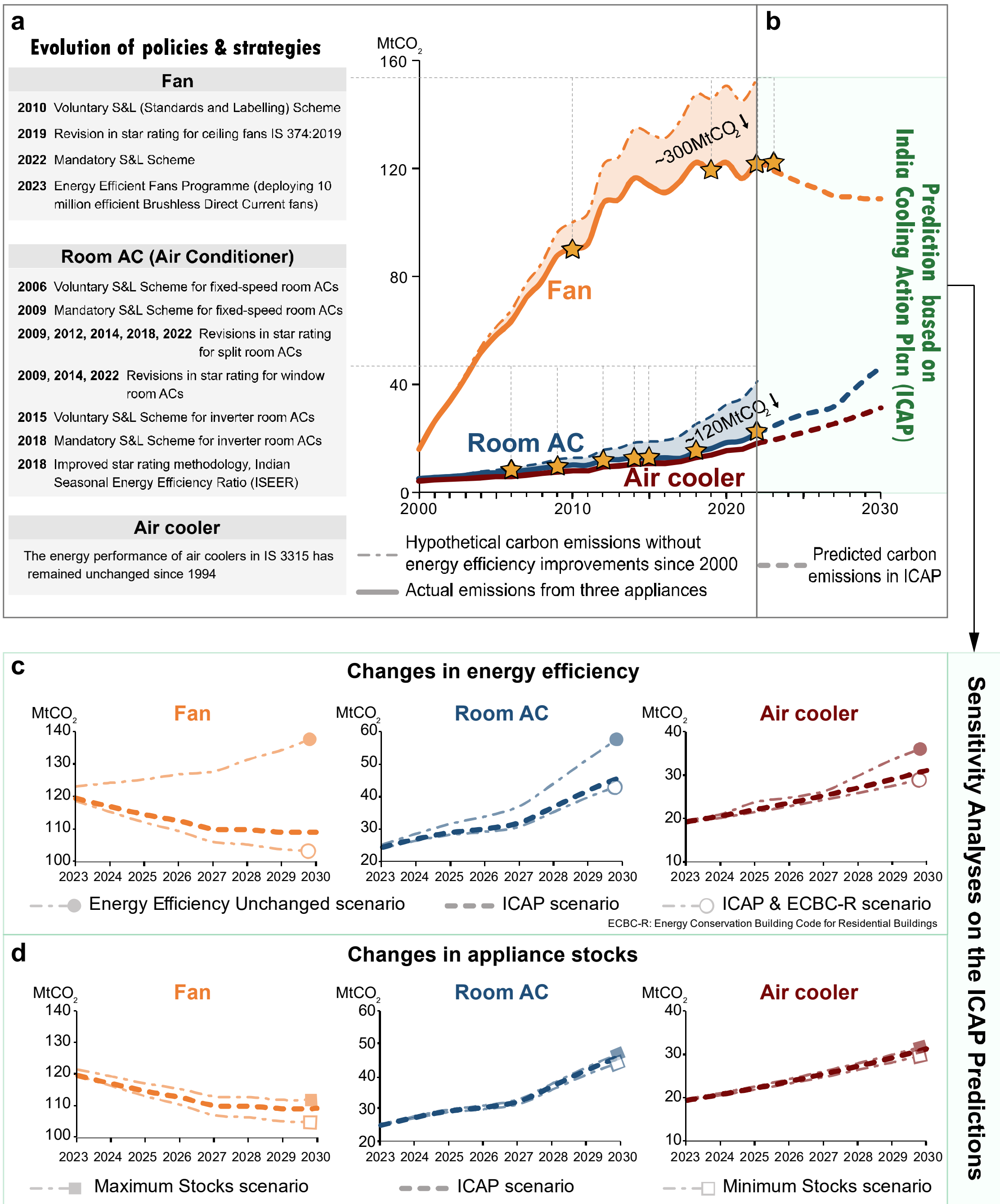

**Fig. 6.** (a) Decarbonization efforts in improving the energy efficiency of fans, room ACs, and air coolers over the past two decades; (b) Predicted carbon emissions through 2030 based on the ICAP, with sensitivity analyses of (c) changes in energy efficiency and (d) changes in appliance stocks. Note: ECBC-R refers to the Energy Conservation Building Code for Residential Buildings, detailed in Section 4.3. Descriptions of the scenarios in the sensitivity analyses can be found in Appendix C.

It is important to note that the predictions in Fig. 6 b were derived by inputting ICAP's assumptions regarding the energy efficiency and stock levels of these three key appliances into the DREAM model. Sensitivity analyses were conducted for the two assumptions mentioned above, as shown in Fig. 6 c and d. For a detailed explanation of the ICAP assumptions and sensitivity analyses, please refer to Appendix C.

### *4.3. Building efficiency pathways for decarbonizing residential space cooling in India*

In response to increasing temperature, nonrefrigerant cooling appliances such as fans and air coolers can provide only limited relief to households during the summer. However, as predicted by the ICAP, even by 2038, a significant portion of households in India may continue to face challenges in affording refrigerant cooling appliances such as room ACs. To address the urgent need for residential space cooling in a widely affordable manner while simultaneously striving to achieve the country's ambitious target of carbon neutrality by 2070, it is essential to prioritize the widespread promotion of climate-responsive and sustainably designed residential spaces [26]. This involves controlling indoor temperatures through passive cooling methods, such as enhancing the energy-efficient design of building envelopes [59], to maintain a comfortable thermal range.

India's initial energy efficiency code for buildings, the Energy Conservation Building Code (ECBC), was introduced in 2007 and underwent a revision in 2017 [60]. Primarily applied to new commercial buildings, its implementation remains voluntary at the national level, providing states with the autonomy to decide whether to revise or mandate it. In 2018, the launch of the energy conservation code for residential buildings, ECBC-R (also referred to as Eco Niwas Samhita) [61], marked a significant milestone, aiming for a 25% reduction in energy consumption compared to that of conventional residential structures, as shown in Fig. 6. Initially implemented as a pilot program in five states—Delhi, Uttar Pradesh, Punjab, Maharashtra, and Karnataka—this code transitioned into a nationwide mandate following the amendment of the Energy Conservation (Amendment) Act, 2022. Applicable to all residential buildings and the residential portions of "mixed land-use construction projects", both built on a plot area of ≥ 500 $m^2$, the code was meticulously crafted to consider the diverse climate zones of India, including the hot dry, warm humid, temperate, composite, and cold zones. This ensures that the energy efficiency measures prescribed are appropriate for each specific climate zone.

According to the anticipation outlined by the Global Buildings Performance Network [15], India's Eco Niwas Samhita will undergo continuous revision and enhancement in the future. For example, ECBC has introduced ECBC+ with savings of 35% and Super ECBC with savings of 50%. Under the moderate pathway, the overall energy efficiency target of new buildings in India is projected to reach 40% by 2050, and under the aggressive pathway, this percentage can further increase to 50%. However, considering India's rapid urbanization and socioeconomic development, it is anticipated that its residential building stock will more than double that of 2022 by 2050, with the total area of new and renovated residential buildings in the next 30 years reaching 40-50 billion $m^2$. In such circumstances, achieving India's 2070 carbon neutrality target remains challenging based on the currently predicted aggressive pathway. This underscores the urgency for India to establish a systematic and viable residential energy efficiency design standard system in the coming decades. This endeavor aims to gradually enhance the energy efficiency targets of new and renovated buildings, striving to achieve greater energy savings and carbon reductions.

As a neighboring developing country, China's experience can provide a valuable reference for shaping the future trajectory of energy efficiency policies and programs in residential buildings in India. As shown in Fig. 7, over the past four decades, China has methodically elevated the design standards for energy efficiency in residential buildings, establishing a four-stage progression with energy efficiency targets of 30%, 50%, 65%, and 75% (compared to conventional residential buildings in the 1980s) [62]. These standards are tailored for different climate zones, including cold, hot summer and cold winter, hot summer and warm winter, and moderate zones. China's initial "Design Standard for Energy Efficiency of Residential Buildings" was introduced in 1986, with a primary focus on cold zones where central heating has been implemented since the 1960s, requiring higher standards for energy-efficient design in building envelopes. Aggressive efforts to enhance energy efficiency in this zone led to the establishment of a 75% target in 2018. Furthermore, other climate zones in China, including hot summers and cold winters, hot summers and warm winters, and moderate climates, all followed a moderate pathway from 2000 to 2022, progressing from a target of 30% to 65%. It is noteworthy that these standards are mandatory, requiring adherence to the latest standards for both new construction buildings and newly renovated structures. Additionally, each province has the autonomy to enhance regulatory requirements based on its specific conditions [63]. For instance, Beijing has emerged as a pioneer in the country, raising the energy efficiency target for residential buildings from 75% to over 80% [64]. In the future, China is progressively moving toward the construction of ultralow-energy, nearly zero-energy, and zero-energy buildings. To support this transition, in 2019, China introduced a voluntary standard titled "Technical Standard for Nearly Zero Energy Buildings" (GB/T 51350-2019), outlining the design requirements for energy efficiency performance in ultralow-energy, nearly zero-energy, and zero-energy buildings. Over the next 30 years, it is anticipated that new buildings in cold zones will prioritize achieving ultralow energy standards (i.e., over the energy efficiency target of 85%) through an aggressive pathway, while other zones may progress along a moderate pathway. Eventually, all zones are projected to meet nearly zero energy standards (i.e., over the energy efficiency target of 90%) by 2050 [65].

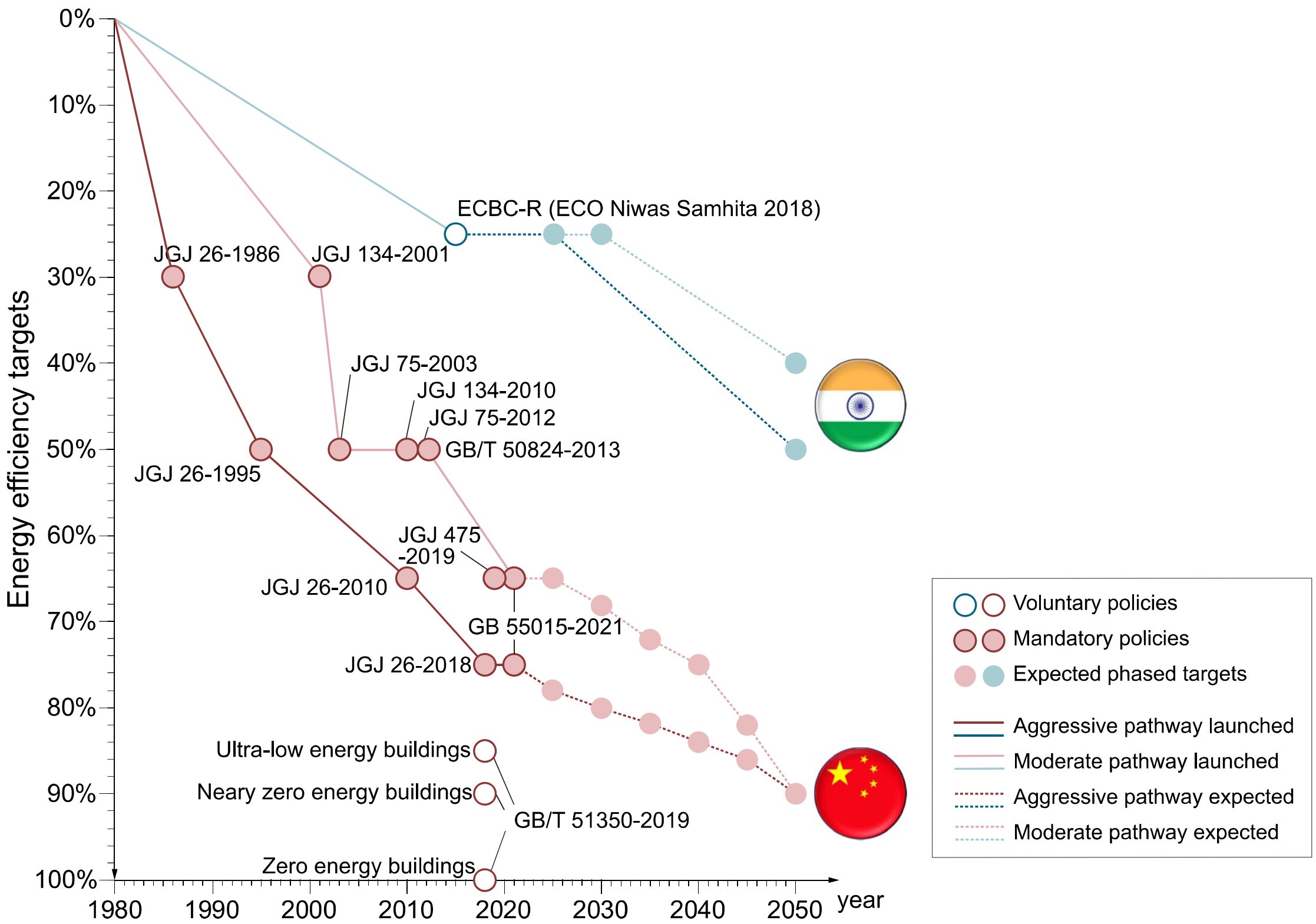

**Fig. 7.** Long-term strategy pathways of residential building design standards for energy efficiency in India and China.

Overall, the study assessed the effectiveness of initiated residential space cooling policies, seeking optimal paths to expedite future decarbonization—addressing Question 3 outlined in Section 1.

## 5. Conclusion

This study undertook a state-level analysis of historical carbon emissions and decarbonization efforts concerning residential space cooling in India. The comprehensive analysis encompassed both urban and rural households across all 33 states and territories of India from 2000 to 2022. Initially, a microlevel investigation of residential space cooling activities was conducted using the DREAM approach, examining the historical evolution of the corresponding carbon intensity. Subsequently, a macrolevel exploration based on the DSD method delved into the societal, economic, climatic, and technological factors influencing changes in carbon intensity for India's residential space cooling. Furthermore, the historical performance of Indian states in decarbonizing urban and rural residential space cooling across four emission scales was examined. Finally, the effectiveness of initiated residential space cooling policies was discussed to identify optimal paths for expediting future decarbonization efforts. The main findings are summarized briefly below.

### *5.1. Key findings*

- **The carbon intensity for space cooling in Indian households increased significantly, with an average annual growth rate of 6.4%, resulting in an average carbon intensity of 513.8 kg$CO_2$ per household in 2022.** The NSDP per capita, representing income, emerged as the primary positive contributor, with a contribution rate of 196.5% to the increase in carbon intensity over the past 22 years. Conversely, the emission factors were the most negative contributors (-41.5%) during this period. When examining the individual performance of each state, Sikkim had the highest average annual growth rate of 8.7% from 2000 to 2022 but the lowest carbon intensity of residential space cooling (43.3 kg$CO_2$ per household) in 2022. In comparison, Chandigarh had the lowest average annual growth rate, 3.5%, but the highest carbon intensity, 1826.5 kg$CO_2$ per household, in 2022.

- **Fans continued to contribute significantly to residential space cooling carbon intensity, accounting for 70.4% and 82.3% of the total in urban and rural areas, respectively, in 2022. The increase in carbon emissions from space cooling was largely driven by the use of fans.** For urban households, 97.7% of the increase in carbon intensity from 2000 to 2011 was attributed to fans, resulting in a surge of 437.6 $kgCO_2$ per household. However, in the subsequent period from 2011 to 2022, carbon intensity from fans decreased by 49.4 $kgCO_2$ per household, with room ACs emerging as the most positive contributor, responsible for an increase of 43.2 $kgCO_2$ per household. In rural India, fans were the primary driver of the increase in carbon intensity, contributing 80.6% and resulting in a surge of 233.5 $kgCO_2$ per household from 2000 to 2022.
- **From 2001 to 2022, Indian households achieved a total cumulative decarbonization of 206.2 $MtCO_2$ for space cooling, with an efficiency of 8.5%. The contribution of the past decade accounted for more than 70% of this total.** Over the 22-year period, urban households achieved a cumulative decarbonization intensity of 1386.1 $kgCO_2$ per household, while rural households reached 368.8 $kgCO_2$ per household. States with higher decarbonization intensity are primarily located in regions with greater ownership of residential space cooling appliances, including Goa, Chandigarh, Delhi, Andhra Pradesh, Gujarat, Tamil Nadu, Puducherry, and Kerala. During the same period, decarbonization per unit of NSDP for space cooling in urban households amounted to 379.7 $kgCO_2$ per lakh rupee, while rural households recorded 218.4 $kgCO_2$ per lakh rupee. States with greater decarbonization per unit of NSDP for space cooling, such as Bihar, Andhra Pradesh, Rajasthan, and Madhya Pradesh, are concentrated in areas with higher annual average temperatures and lower household incomes.

### *5.2. Upcoming work*

This work endeavors to further explore the optimal pathways for achieving high decarbonization in residential space cooling in India, a major emerging emitter, in accordance with the nation's goal of achieving zero-carbon emissions by 2070. Future related work will address two key issues. First, future decarbonization pathways under various climate scenarios will be explored.

By considering shared socioeconomic pathways and global warming of 1.5-2 °C, dynamic simulations will be developed to analyze future carbon emissions and decarbonization trends from urban and rural residential space cooling in India across multiple scenarios and scales. Second, India's future key decarbonization strategies should be investigated. Drawing insights from policies and technological advancements in space cooling decarbonization worldwide, strategies tailored to India's national conditions and social development will be identified. Additionally, cost–benefit analyses will be conducted to determine the optimal combination of decarbonization strategies.

## Appendix

Please find the appendix in the supplementary materials (e-component).

## Acknowledgments

First author appreciates the National Planning Office of Philosophy and Social Science Foundation (24BJY129).

## Author contributions

M.M., N.Z., and C.M. constructed the analytical framework. R.Y. and M.M. developed the original model. R.Y., M.M., and N.Z. performed the simulations and analyzed the data. All authors contributed to the analyses and writing.

## Declaration of interests

The authors declare no competing interests.